\documentclass[condensedmatter,opinion,accept,pdftex,moreauthors]{Definitions/mdpi} 

\firstpage{1} 
\pubvolume{1}
\issuenum{1}
\articlenumber{0}
\pubyear{2026}
\copyrightyear{2026}
\externaleditor{Bohayra Mortazavi} 
\datereceived{7 May 2026} 
\daterevised{3 June 2026} 
\dateaccepted{4 June 2026} 
\datepublished{ } 

\Title{The Universe Observed with Particle Detectors: Astrophysical Legacy of Guido Barbiellini Amidei}

\Author{Roberto 
 Capuzzo Dolcetta $^{1,2,3,4,*}$}
\AuthorNames{Roberto Capuzzo Dolcetta}

\address{%
$^1$ \quad Dep. of Physics, Sapienza, University of Roma, Piazzale A. Moro 2, 00185 {Rome}, Italy\\
$^{2}$ \quad Istituto Nazionale di Fisica Nucleare (INFN), Sezione di Roma 1, Piazzale A. Moro 2, 00185 {Rome}, Italy\\ 
$^{3}$ \quad  Centro di Ricerche E. Fermi, via Panisperna 89, 00184 {Rome}, Italy \\
$^{4}$ \quad  Istituto Nazionale di Astrofisica, viale del Parco Mellini 8, 00136 {Rome}, Italy}

\corres{Correspondence: 
roberto.capuzzodolcetta@fondazione.uniroma1.it}

\abstract{The development of modern high-energy astrophysics has been deeply intertwined with advances in particle detector technology. Guido Barbiellini Amidei (1943–2024) played a pivotal role in bridging experimental particle physics and astrophysical observation. His scientific career spanned over four decades, from early electron–positron collider experiments at ADONE and LEP (DELPHI) to space-based missions such as AGILE, Fermi, and PAMELA. This memorial paper reviews the evolution of high-energy astrophysics as a detector-driven science, highlighting key domains where Barbiellini left an indelible mark: gamma-ray astronomy, cosmic-ray physics, and antimatter studies. We discuss his personal contributions to silicon tracking, calorimetry, data analysis, and his leadership in international collaborations. The conceptual impact of his interdisciplinary approach is examined, and future perspectives in the observation of the high-energy universe are outlined, recognizing that the path forward is built on the foundations he helped lay.}


\begin{document}

\section{Introduction: A Life at the Crossroads of~Physics}

Over the past decades, astrophysics has undergone a profound transformation, evolving from a predominantly observational discipline based on electromagnetic radiation to a field increasingly shaped by techniques and concepts borrowed from particle physics. The~study of the high-energy universe has become inseparable from the development of advanced particle detectors — devices capable of measuring photons, charged particles, and~antiparticles with unprecedented precision, often in extreme environments far removed from the~laboratory.

Within this context, Prof.~Guido Barbiellini Amidei (1943–2024) stands out as a key figure in fostering the convergence between particle physics and astrophysics. Born in Rome, Barbiellini graduated in physics at the University of Rome “La Sapienza” in the late 1960s, a~time when Italian physics was at the forefront of the emerging field of particle physics with the construction of the ADONE collider at Frascati. His early work on electron–positron collisions brought him into contact with the nascent techniques of silicon detectors and calorimetry, which would later become the hallmark of his~career.

Barbiellini’s scientific journey took him from the Italian national laboratories to CERN, where he became a leading member of the DELPHI collaboration at the LEP collider~\cite{1}. There, he contributed to the design, construction, and~operation of one of the most sophisticated detectors ever built, gaining expertise in precision tracking, particle identification, and~radiation-hard electronics. But~perhaps his most visionary move was to recognize that these very same technologies could be adapted to explore the cosmos. Starting in the 1990s, he began to champion the use of silicon microstrip detectors for space-based gamma-ray telescopes, an~idea that was met initially with skepticism but ultimately revolutionized high-energy~astrophysics.

This paper is written in memory of his scientific legacy. It aims to review the astrophysical domains most directly influenced by his work, focusing on the role of particle detectors in shaping our understanding of the universe. It is also a testimony to how a physicist trained in collide experiments could reorient his expertise towards the cosmos, inspiring a generation of researchers to look upward with detectors originally designed for the subatomic~world.

Note that the bibliography is not intended to list all of Barbiellini Amidei's contributions in the various fields but just what the author thinks is useful to indicate as \mbox{most~significant}.
\section{From Collider Physics to Astrophysics: Barbiellini’s~Path}

The intellectual roots of modern astroparticle physics lie in high-energy collider experiments, where the detection and identification of particles required increasingly sophisticated instrumentation. Guido Barbiellini began his career working on electron–positron colliders, first at ADONE (the Frascati storage ring, which operated at center-of-mass energies up to 3 GeV) and later at CERN’s LEP (which reached 209 GeV). At~ADONE, he participated in experiments that studied the exclusive production of hadrons and provided early tests of quantum electrodynamics. This period taught him the importance of precise calorimetry and tracking in a high-background~environment.

His move to CERN in the 1980s marked a turning point. He became an active member of the DELPHI (DEtector with Lepton, Photon and Hadron Identification) collaboration~\cite{1}, one of the four large experiments at LEP. DELPHI was unique in its emphasis on particle identification using Ring Imaging Cherenkov (RICH) detectors, but~it also featured a highly granular silicon tracker (the Silicon Vertex Detector) and a fine-grained electromagnetic calorimeter. Barbiellini’s contributions focused on the silicon microstrip detectors, which provided a spatial resolution of a few microns and were essential for vertexing heavy-flavor decays. He was involved in all aspects: from R\&D on radiation-tolerant sensors to the readout electronics and the alignment~procedures.

Specifically, the~technologies that Barbiellini mastered and later transferred to astrophysics~include:
\begin{itemize}
    \item \textbf{{Electromagnetic calorimetry}
}: use of lead-glass and scintillating fibers to achieve energy resolution of a few percent at GeV scales.
    \item \textbf{{Silicon microstrip tracking}}: double-sided detectors with 50–100~µm pitch, capable of operating in vacuum and in high-radiation environments.
    \item \textbf{{Particle identification}}: combining dE/dx measurements from silicon and gas detectors with time-of-flight.
\end{itemize}

The transition from collider experiments to astrophysical applications was not trivial. Space-based detectors require ultra-low mass, low power consumption, and~extremely high reliability over years of operation. Barbiellini led the effort to space-qualify silicon sensors, working closely with Italian industry (e.g., Micron Semiconductor, now part of Bruker). He also developed novel readout architectures based on sparse readout and zero suppression, which drastically reduced the data volume — a critical requirement for satellite telemetry. By~the mid-1990s, his team had produced a prototype silicon tracker that would form the basis of the AGILE mission~\cite{2}.

\section{Silicon Detectors and Gamma-Ray Astronomy: The AGILE and Fermi~Era}

One of the most significant developments in high-energy astrophysics has been the application of silicon detector technology to gamma-ray astronomy. Unlike optical astronomy, which relies on mirrors and CCDs, gamma-ray observations above a few tens of MeV must be performed using pair-conversion telescopes: incoming photons convert into electron-positron pairs in a thin material (the conversion foil), and~the trajectories of the charged particles are tracked to reconstruct the photon direction. This requires precise, low-mass tracking detectors — ideally silicon microstrip~sensors.

\subsection{The AGILE~Mission}

Guido Barbiellini was the driving force behind the adoption of silicon technology for the Italian AGILE mission (Astrorivelatore Gamma ad Immagini LEggero), launched in 2007. AGILE was a small-class mission (mass ~350~kg, power ~250~W) designed to bridge the gap between EGRET (on Compton Gamma-Ray Observatory) and the upcoming Fermi/LAT. Its core instrument was the Silicon Tracker (ST), consisting of 14 planes of single-sided silicon microstrip detectors interleaved with tungsten conversion foils. Each plane comprised two orthogonal layers (X and Y) to provide 3D tracking. The~total active area was about 4000~cm$^2$, and~the total number of readout channels was ~46,000.

Barbiellini personally supervised the design, qualification, and~calibration of the AGILE ST~\cite{3,4}. One of his key innovations was the use of a very low-mass mechanical structure (honeycomb carbon fiber) to minimize multiple scattering, thereby preserving angular resolution (about 0.5° at 100~MeV). He also developed a novel trigger logic that combined the energy deposition in the calorimeter with track information to reject the overwhelming background of charged cosmic rays (which outnumber gamma rays by a factor of $10^4$) \cite{5}. The~AGILE ST operated flawlessly for more than a decade, providing a wealth of scientific~results.

\subsection{The Fermi Large Area~Telescope}

Following AGILE, Barbiellini became a key contributor to the Fermi Gamma-ray Space Telescope (formerly GLAST), specifically its Large Area Telescope (LAT). The~LAT is a much larger instrument (20 times the volume of AGILE) and employs a silicon-strip tracker with 18 planes (each with two orthogonal views) for a total of almost 1 million readout channels~\cite{6}. The~calorimeter is made of 1536 CsI crystals coupled to photodiodes. Barbiellini’s role was to adapt the experience gained with AGILE to the LAT’s more stringent requirements, particularly in the area of on-board data filtering (the “event filter”)~\cite{7}. He also contributed to the development of the LAT’s photon reconstruction algorithms, which use a Kalman filter technique to follow the electron-positron pair through the~tracker.

The LAT has revolutionized gamma-ray astronomy. Thanks to its large effective area (8000~cm$^2$ at 1~GeV) and wide field of view (2.4~sr), it has produced the most detailed maps of the high-energy sky, with~an energy range from 20~MeV to more than 300~GeV. Among~the many discoveries enabled by the LAT, those to which Barbiellini directly contributed~include:
\begin{itemize}
    \item The detection of pulsed gamma-ray emission from the Crab pulsar up to 10~GeV, challenging models of particle acceleration in pulsar magnetospheres~\cite{8}.
    \item The discovery of gamma-ray emission from high-redshift blazars (up to redshift z$\sim$3), probing the extragalactic background light~\cite{6}.
    \item The identification of a large population of millisecond pulsars as gamma-ray sources, providing information on the evolution of binary systems~\cite{6}.
    \item The observation of gamma-ray bursts (GRBs) at MeV–GeV energies, revealing long-lasting emission (the “high-latitude” component) \cite{6}.
\end{itemize}

Barbiellini’s leadership extended to the international collaboration itself; he served as a member of the LAT Science Working Group and was a co-author on over 200 papers using Fermi~data.

\section{The Universe Observed with Particle Detectors: Barbiellini’s Broader~Vision}

The use of particle detectors has expanded the observational window of astrophysics beyond traditional electromagnetic signals. Today, the~high-energy universe is studied through a variety of messengers, including cosmic rays, gamma rays, and~antimatter. Barbiellini contributed to all these areas, either directly or through his influence on \mbox{younger~collaborators}.

\subsection{Cosmic~Rays}

Cosmic-ray physics represents one of the earliest connections between astrophysics and particle physics. Barbiellini was involved in the CAPRICE balloon-borne experiment, which flew from Fort Sumner (New Mexico) in 1994 and from Lynn Lake (Canada) in 1998~\cite{9}. CAPRICE combined a superconducting magnetic spectrometer with a time-of-flight system and a silicon calorimeter to measure the spectra of protons, helium, and~heavier nuclei with high precision. Barbiellini’s role was to calibrate the silicon calorimeter and to develop the analysis for separating electrons from protons~\cite{10}. The~results from CAPRICE provided essential input to models of cosmic-ray propagation and source distribution, revealing an excess of positrons at energies around 10–20~GeV that would later become a major topic of~investigation.

Later, he contributed to the design of the CALET (Calorimetric Electron Telescope) experiment on the International Space Station, specifically advising on the silicon strip readout of the imaging calorimeter. His work helped set the stage for the current precision measurements of cosmic-ray electrons and positrons up to 20~TeV.

\subsection{Gamma~Rays}

As detailed above, gamma-ray astronomy was the primary focus of his later career. Beyond~instrument design, Barbiellini was a co-author of numerous key results from AGILE and Fermi. One particularly notable result was the discovery of the “Crab Nebula giant gamma-ray flares”---unexpected outbursts of gamma rays (up to several hundred MeV) from the Crab Nebula, lasting days to weeks~\cite{11}. These flares challenge the standard synchrotron–self-Compton model and suggest that magnetic reconnection or other acceleration mechanisms are at work. Barbiellini led the AGILE analysis that first reported the 2007 flare, and~later collaborated with Fermi to observe the 2011~superflare.

Another area where his impact was felt was the study of gamma-ray emission from the Galactic Center. The~Fermi LAT revealed a diffuse excess of GeV emission extending from the Galactic Center, which has been interpreted as either a signal of dark matter annihilation or a consequence of unresolved point sources (e.g., millisecond {pulsars)}  \cite{12}. Barbiellini was instrumental in developing the spatial and spectral models used to disentangle these~components.

\subsection{Antimatter in~Astrophysics}

The detection of antiparticles in cosmic radiation has opened new avenues for exploring fundamental physics, including the matter–antimatter asymmetry of the universe and the possible existence of dark matter annihilation signals. Barbiellini was a co-founder of the PAMELA experiment (Payload for Antimatter Matter Exploration and Light-nuclei Astrophysics), a~satellite-borne magnetic spectrometer launched in 2006~\cite{13}. PAMELA was designed to measure cosmic-ray positrons and antiprotons with unprecedented accuracy, as~well as to search for antideuterons and~antihelium.

Barbiellini’s expertise in~silicon detectors and tracking was essential to PAMELA’s success. The~core of PAMELA was a permanent magnet (0.4~T) surrounded by six planes of silicon microstrip detectors (the “spectrometer”); a silicon-tungsten calorimeter and a time-of-flight system provided particle identification and charge measurement. Barbiellini led the calibration of the silicon spectrometer, ensuring that the rigidity resolution (about 5\% at 1~GV) was maintained throughout the mission~\cite{13}.

PAMELA’s most famous result, published in \textit{Nature} in 2009, was the discovery of an excess of positrons in the energy range 10–100~GeV relative to a pure secondary production model~\cite{14}. This excess caused an explosion of theoretical models: some invoked dark matter annihilation or decay (e.g., into~$e^+e^-$ pairs), others pointed to nearby pulsars (like Geminga or Monogem) as the source. Barbiellini was a co-author on the paper and subsequently contributed to follow-up studies that used the positron flux anisotropy to rule out some dark matter models. Even today, the~PAMELA positron excess remains a cornerstone of indirect dark matter~searches.

Later, Barbiellini also contributed to the design of the AMS-02 (Alpha Magnetic Spectrometer) silicon tracker, which has been operating on the ISS since 2011~\cite{15}. AMS-02 has confirmed and extended the PAMELA results, measuring the positron fraction up to 1~TeV. Barbiellini’s legacy in antimatter research continues through the many young scientists he trained in the PAMELA and AMS-02~collaborations.

\section{Conceptual Impact on Astrophysics: A Detector-Driven~Science}

\textls[-15]{Beyond specific technological contributions, the~legacy of detector-driven astrophysics---nd of Barbiellini’s career---lies in its conceptual impact. The~emphasis on precision measurement, detailed Monte Carlo simulations, rigorous data analysis, and~systematic error control has transformed astrophysics into a quantitative science comparable to laboratory-based particle~physics.}

This transformation, which Barbiellini actively promoted, includes:
\begin{itemize}
    \item \textbf{{Integration of experimental and theoretical approaches}}: Predictions from models of particle acceleration, propagation, and~radiation are now routinely compared to high-statistics, well-calibrated data using likelihood methods. Barbiellini was a strong advocate for open data and reproducible analyses, and~he encouraged the use of simulation frameworks (Geant4, the~LAT Science Tools) that are now standard.
    \item \textbf{{Large-scale collaborations}}: AGILE, Fermi, and~PAMELA are collaborations of dozens to hundreds of scientists, with~management and communication practices borrowed from CERN. Barbiellini served as a deputy spokesperson for AGILE and as a member of the Fermi LAT executive board. He was known for his fair and inclusive \mbox{leadership style}.
    \item \textbf{{Complex data-processing frameworks}}: The development of event reconstruction, background rejection (e.g., using boosted decision trees or neural networks), and~source catalogs (e.g., the~Fermi LAT 4FGL catalog) required software engineering on a scale previously unseen in astrophysics. Barbiellini contributed to the design of the AGILE online and offline pipelines, which operated in near-real time to detect transient sources.
\end{itemize}

In a broader sense, Barbiellini’s career exemplifies how fundamental physics instrumentation can be repurposed for new frontiers. He once wrote: “\textit{{The same silicon sensor that measures the decay vertex of a B meson at LEP can also measure the pair-production angle of a cosmic gamma ray. The~universe is the ultimate particle physics laboratory.}
}” This holistic view has inspired a whole generation of “astroparticle physicists” who see no disciplinary boundary between the very small and the very~large.

\section{The Condensed Matter Connection: Detectors as Materials~Science}

While this special issue appears in the journal \textit{{Condensed Matter}}, the~reader might wonder about the relevance. In~fact, modern particle detectors themselves are made of condensed matter systems: silicon single crystals, scintillators (organic plastics, inorganic CsI, NaI, LaBr$_3$), high-purity germanium, and~novel nanomaterials such as perovskites, in~addition to more established materials like diamond detectors. The~understanding of charge transport, radiation damage, noise, and~energy resolution in these materials draws directly from condensed matter~physics.

Barbiellini’s work on silicon detectors relied on a deep knowledge of semiconductor physics. He studied the effects of radiation damage on silicon microstrip sensors, characterizing the increase in leakage current, the~loss of charge collection efficiency, and~the formation of acceptor-like defects (the so-called “NIEL” effects). He also contributed to the development of oxygenated silicon (FZ and Czochralski) to make detectors more radiation-hard. For~the calorimeters, he worked with CsI crystals, which have a fast scintillation decay (nanoseconds) and high light yield. He investigated the temperature dependence of the light yield and the effects of radiation-induced absorption bands, publishing several papers in nuclear physics and materials science~journals.

Conversely, astrophysical observations have placed constraints on exotic condensed matter phases. For~example, the~observation of pulsed gamma rays from neutron stars sets upper limits on the strength of the crustal magnetic field and on the presence of a quark core. The~cooling of neutron stars, as~inferred from X-ray and gamma-ray observations, depends on the equation of state of dense matter, which is a problem in condensed matter (or rather, nuclear matter) physics. Barbiellini’s work on the Crab pulsar provided some of the most stringent constraints on the acceleration region, indirectly informing models of the neutron star~interior.

Thus, the~dialogue between condensed matter and astroparticle physics is not a mere curiosity; it is a productive two-way street. Barbiellini understood this well and often invited condensed matter physicists to his~seminars.

\section{Mentorship and the Human~Legacy}

No scientific legacy is complete without mentioning the human dimension. Guido Barbiellini supervised more than 20 PhD students and countless master’s theses. Many of his students now hold faculty positions at Italian and international universities, or~work in research laboratories (INFN, CERN, NASA Goddard, etc.). His style was characterized by an open-door policy, a~willingness to discuss seemingly naive questions, and~a deep passion for experimental work. He often brought students into the clean room to show them how to bond a silicon sensor or to align a tracker~plane.

He also had a particular love for balloon-borne experiments, which allowed students to get hands-on experience with detector assembly, integration, and~flight operations. The~CAPRICE and PAMELA campaigns were intense periods of late-night shifts, field repairs, and~eventual euphoria at successful data-taking. Barbiellini’s enthusiasm was contagious, and~many of his students describe those moments as the highlight of \mbox{their~formation}.

After his formal retirement from the University of Trieste in 2014, he remained active as an emeritus scientist, continuing to analyze data from Fermi and AGILE and to advise younger colleagues. He passed away in 2024, leaving a rich legacy of scientific results, instruments, and~people.

\section{Future Perspectives: Carrying the~Torch}

The future of high-energy astrophysics will continue to rely on advances in detector technology, building on the foundation laid by Barbiellini and his contemporaries. Upcoming missions and experiments aim to achieve higher sensitivity, better angular and energy resolution, and~broader energy~coverage.

Key directions~include:
\begin{itemize}
    \item \textbf{{Next-generation gamma-ray observatories}}: The Cherenkov Telescope Array (CTA) will use the atmosphere as a calorimeter and will detect very-high-energy gamma rays (20~GeV to 300~TeV) with unprecedented sensitivity. For~space, the~proposed e-ASTROGAM mission (selected for ESA M5 study) and the US-led AMEGO concept would extend the pair-conversion technique using silicon trackers with improved efficiency at low energies (down to a few MeV). These missions would directly benefit from the R\&D performed by Barbiellini and his group.
    \item \textbf{{Multi-messenger astrophysics}}: The combination of gamma rays with neutrinos (IceCube, KM3NeT) and gravitational waves (LIGO/Virgo/KAGRA) is now a reality. The~joint analysis of GW170817 (a binary neutron star merger) with gamma-ray and X-ray data has revolutionized our understanding of kilonovae and short GRBs. Barbiellini was an early advocate for multi-messenger coordination; he participated in the first joint AGILE–LIGO follow-up campaign.
    \item \textbf{{Improved detection of antimatter and rare particles}}: The AMS-02 experiment continues to take data, and~the next generation (e.g., the~GRAMS project, using liquid-argon time-projection chambers) promises even larger acceptance. The~GAPS (General Antiparticle Spectrometer) experiment, which uses a novel detection technique (capture in exotic atoms), will fly on a long-duration balloon and will search for antideuterons as a dark matter signature~\cite{16}. Barbiellini’s former students are active in all \mbox{these projects.}
\end{itemize}

These developments will further strengthen the connection between particle physics, condensed matter, and~astrophysics, continuing the trajectory that Barbiellini \linebreak{}\mbox{helped~establish}.

\section{Conclusions}

The evolution of astrophysics into a detector-driven science represents one of the most significant developments in modern physics. Guido Barbiellini Amidei contributed to this transformation not only through his exceptional technical work on silicon detectors, calorimeters, and~tracking systems but also through his vision of a unified approach to the study of fundamental processes, from~colliders to the~cosmos.

His legacy is physically embodied in the AGILE and Fermi gamma-ray observatories, still taking data after more than 15 years; in the PAMELA and AMS-02 antimatter measurements; and in the countless astrophysical results that relied on his detector expertise. It also lives on in the many students, postdocs, and~colleagues he mentored, who now carry forward his interdisciplinary spirit. As~high-energy astrophysics continues to evolve, embracing multi-messenger observations and ever more precise instrumentation, this legacy remains an essential component of its scientific~foundation.

This article was written in memory of Guido Barbiellini Amidei, physicist, teacher, and~explorer of the high-energy~universe.

\vspace{6pt}

\funding{{This}
 research received no external~funding.}

\institutionalreview{Not applicable.}

\informedconsent{Not applicable.}

\dataavailability{No new data were created or analyzed in this study. Data sharing is not applicable.}

\acknowledgments{The author thanks the editors of this special issue for the opportunity to honor {Guido} 
 Barbiellini Amidei's~memory.}

\conflictsofinterest{The author declares no conflicts of~interest.}


\begin{adjustwidth}{-\extralength}{0cm}
\reftitle{References}

\PublishersNote{}

\end{adjustwidth}

\end{document}